\documentclass[conference]{IEEEtran}
\IEEEoverridecommandlockouts
\usepackage{cite}
\usepackage{amsmath,amssymb,amsfonts}
\usepackage{algorithmic}
\usepackage{graphicx}
\usepackage{textcomp}
\usepackage{xcolor}
\usepackage[table]{xcolor}
\usepackage{booktabs}
\usepackage{tabularx}
\usepackage{siunitx}
\usepackage[T1]{fontenc}
\usepackage{booktabs}
\usepackage{amssymb} 
\usepackage[utf8]{inputenc}
\usepackage{ragged2e}
\usepackage{multirow}
\usepackage{array}

\usepackage{url}
\usepackage[hidelinks]{hyperref}
\hypersetup{
    pdfborder={0 0 0}
}

\def\BibTeX{{\rm B\kern-.05em{\sc i\kern-.025em b}\kern-.08em
    T\kern-.1667em\lower.7ex\hbox{E}\kern-.125emX}}
\begin{document}

\title{
When LLMs Slow Down: How Environmental Impacts Mediate University Students’ LLM Usage
}



% ... 

\author{
\IEEEauthorblockN{
\begin{tabular*}{\textwidth}{@{\extracolsep{\fill}} >{\centering\arraybackslash}m{0.3\textwidth} >{\centering\arraybackslash}m{0.3\textwidth} >{\centering\arraybackslash}m{0.3\textwidth} @{}}
Hyeonwook Kim & Xuesi Chen & Alex Cabral
\end{tabular*}}
\IEEEauthorblockA{
\begin{tabular*}{\textwidth}{@{\extracolsep{\fill}} >{\centering\arraybackslash}m{0.3\textwidth} >{\centering\arraybackslash}m{0.3\textwidth} >{\centering\arraybackslash}m{0.3\textwidth} @{}}
\textit{Georgia Institute of Technology} & \textit{Cornell Tech} & \textit{Georgia Institute of Technology} \\
Atlanta, GA, USA & New York, NY, USA & Atlanta, GA, USA \\
hkim3159@gatech.edu & xc562@cornell.edu & acabral30@gatech.edu
\end{tabular*}}
\and
\IEEEauthorblockN{
\begin{tabular*}{\textwidth}{@{\extracolsep{\fill}} >{\centering\arraybackslash}m{0.3\textwidth} >{\centering\arraybackslash}m{0.3\textwidth} >{\centering\arraybackslash}m{0.3\textwidth} @{}}
Cindy Kaiying Lin & Udit Gupta & Josiah Hester
\end{tabular*}}
\IEEEauthorblockA{
\begin{tabular*}{\textwidth}{@{\extracolsep{\fill}} >{\centering\arraybackslash}m{0.3\textwidth} >{\centering\arraybackslash}m{0.3\textwidth} >{\centering\arraybackslash}m{0.3\textwidth} @{}}
\textit{Georgia Institute of Technology} & \textit{Cornell Tech} & \textit{Georgia Institute of Technology} \\
Atlanta, GA, USA & New York, NY, USA & Atlanta, GA, USA \\
clin646@gatech.edu & ugupta@cornell.edu & josiah@gatech.edu
\end{tabular*}}
}

    





\maketitle

\begin{abstract}
Large Language Models (LLMs) are increasingly being embedded into all facets of society, from search to education, industrial, and financial applications. These systems' carbon and water footprints raise important sustainability concerns, particularly with adoption rates exceeding 80\% among university students, despite limited insight into the environmental impacts of individual usage. Eco-feedback interfaces offer a promising approach to encourage more sustainable behaviors, yet their role in shaping LLM users’ sustainability awareness and decision-making remains underexplored. We design and deploy the interface that visualizes latency–carbon trade-offs during live LLM interactions. We study its use with undergraduate computer science students ($N=89$, ages 18–24),  enrolled in a computing ethics course, providing an empirical look at how a technically sophisticated and values-oriented user population responds to sustainability-aware AI interfaces. We found that the likelihood of choosing the eco-feedback system significantly decreased as perceived response latency increased ($p < .001$), while users’ willingness increased when they recognized the carbon-saving impacts ($p < .01$). Also, students with stronger eco-mindedness demonstrated higher baseline willingness to adopt lower-carbon modes and reported increased awareness of the environmental impacts of LLM use, though this effect diminished as latency increased. These results position eco-feedback interfaces as a promising sustainability intervention and highlight their potential as an educational opportunity to promote more sustainable LLM use among university students and beyond.

\end{abstract}

\begin{IEEEkeywords}
Eco-Feedback, Digital Carbon Emission, Sustainability, Human-Computer Interaction, Computer Science
\end{IEEEkeywords}


\section{Introduction}

The environmental footprint of Large Language Models (LLMs) is substantial and growing. Increased LLM use has been linked to rising greenhouse gas (GHG) emissions, water consumption, and rare earth mineral extraction~\cite{luccioni2025efficiency}. Among these impacts, GHG emissions are particularly critical due to their central role in climate change~\cite{IPCC2023_AR6}. Over the past five years, Google and Microsoft reported GHG emission increases of more than 48\% and 29\%, respectively, largely driven by data center energy consumption~\cite{luccioni2025efficiency}. Independent analysis further suggests that actual data center emissions may exceed reported figures by up to 600\%~\cite{Guardian2024_DataCenterEmissions}. This trend is driven by rapidly increasing demand for data center power, with projections indicating that electricity use could nearly double by 2026 due in large part to AI workloads~\cite{KouLimandibhratha2025}. Even individual AI models incur substantial environmental costs; for example, training GPT-3 is estimated to have produced 552 metric tons of CO$_2$ equivalent emissions~\cite{Oliver2024responsible}.
However, many users remain unaware of the direct sustainability impacts of their individual LLM use, making it unclear how such information would shape their interaction choices. 

Eco-feedback is a well-established strategy in Human–Computer Interaction (HCI) that utilizes visual information about sustainability impacts to change user behaviors~\cite{Froehlich2010_DesignEcoFeedback, Sohn2015effects}. However, eco-feedback design and application in LLM-based systems remain largely unexplored~\cite{GrosserOlivares2025, MorrisonEtAl2025}. This approach may be particularly relevant in educational contexts, where values, habits, and professional norms around technology use are actively formed. University students represent one of the most frequent and active user groups of LLMs, with adoption rates exceeding 80\%, yet often lack insight into the environmental impacts of individual usage~\cite{programs_ai_stats_2025, umd_genai_survey_2024, reyes_ai_adoption_2024}. 

At the same time, young adults aged 18–24 are among the most willing to engage in pro-environmental practices and activism~\cite{Latkin2021correlates, Gomes2023willingness}, with surveys indicating that approximately 65\% have used generative AI tools~\cite{Young2024role}. Within this population, computer science (CS) students occupy a unique position as both high-frequency users and future creators of AI systems. In university settings, LLMs are increasingly embedded in learning, coursework, and programming tasks. Eco-feedback interfaces therefore offer the potential to reduce immediate environmental impact if lower-impact modes are adopted over time. They also create an opportunity to support learning about LLM trade-offs and to encourage longer-term shifts in usage behavior, such as tolerating latency, refining prompting strategies, or reducing unnecessary use.

This motivates the following research questions:
\begin{itemize}
    \item \textbf{RQ1:} How do eco-feedback design factors, including environmental impact representations and response latency, shape users’ eco-mode preferences and perceived usability in LLM interactions?
    \item \textbf{RQ2:} How do users’ eco-mindedness and interpretations of eco-feedback influence their engagement with and willingness to adopt eco modes under latency conditions?
\end{itemize}




To address these questions, we designed an eco-feedback LLM interface and evaluated it with 89 university students in the US. Central to our approach is the use of service latency as a design lever for reducing carbon emissions, building on prior work showing that users are often willing to tolerate slower system responses in exchange for sustainability benefits~\cite{Kim2025slower, Ting2025impact}. Drawing on environmental psychology and HCI research, the interface is grounded in two design principles. First, it leverages the motivational power of visualization: making environmental impacts visible supports the internalization of pro-environmental goals and meaningful action~\cite{Boomsma2016imagining, Salazar2022testing}. Second, it engages users’ personal environmental norms, or eco-mind, recognizing moral obligation as a strong predictor of pro-environmental behavior~\cite{Froehlich2010_DesignEcoFeedback, Sohn2015effects}. Guided by these considerations, we examine how eco-feedback in LLM interactions relates to sustainability-oriented use under different practical conditions and how interacting with such an interface shapes users’ understanding of the environmental implications of everyday LLM use.

Our findings show that the sustainability benefits of eco-feedback in LLM use are constrained by users’ tolerance for latency. Eco-mode selection declined sharply as latency increased, dropping from approximately 45\% of interactions in eco mode~1 to below 5\% in eco mode~5. Mixed-effects models confirm that higher perceived latency significantly reduced eco-mode preference ($p < .001$), while recognition of carbon-saving impact significantly increased eco-mode selection under low-latency conditions ($p < .001$). Personal environmental norms played an important but limited role: participants with stronger eco-mindedness showed higher initial eco-mode preference, but this effect diminished rapidly as response delays increased, and even environmentally motivated users abandoned eco modes when latency became disruptive.

The results also point to sustainability and educational potential under usability-compatible conditions. Participants reported increased awareness of the environmental impacts of everyday LLM use, including the relationship between latency, computation, and emissions, and 
reflected on how this awareness could shape their interaction choices. In addition, more concrete and outcome-oriented eco-feedback visualizations were rated more favorably than abstract metrics, supporting user engagement and understanding. Through this work, we contribute to exploring user-acceptable eco-feedback interfaces, their relationship to lower-carbon LLM use, and how interaction with such systems surfaces gaps in understanding the environmental impact of everyday AI use.

\section{Related Works}




\subsection{Eco-feedback Interfaces} 
Within Sustainable HCI~\cite{blevis2007sustainable,disalvo2010mapping,hansson2021decade}, researchers have increasingly developed eco-feedback systems to provide information to users with the goal of reducing environmental impact~\cite{Froehlich2010_DesignEcoFeedback,mccalley1998computer}. Many of these works rely on gamification, motivational techniques, and informational display elements to change individual and group behaviors for reduced energy consumption and greenhouse gas emissions~\cite{guizzardi2025enhancing,di2017vehicle,Chalal2022}. Common display techniques utilize green coloring, eco-related icons such as leaves, cost savings for energy usage, and relatable information about greenhouse gas emissions such as miles driven or light bulb usage~\cite{guizzardi2025enhancing,shamma2022ev,bao2016eco}. Prior works have found that such feedback can influence decision-making in driving patterns, transportation mode selection, and energy consumption in buildings~\cite{di2017vehicle,ma2018longitudinal,froehlich2009ubigreen}. Furthermore, researchers have found that eco-feedback designs incorporating social norms and emotionally evocative information can encourage sustainable user behavior~\cite{starke2020little,bao2016eco}. 

Yet eco-feedback interfaces have also faced critiques~\cite{hansson2021decade}. Researchers have noted that such ``persuasive sustainability'' offers just a narrow view of sustainability and places too much weight on the assumption that users will make sustainable choices if provided enough information~\cite{brynjarsdottir2012sustainably,hansson2021decade}. Others have suggested that designers should not push to engineer behavior change but rather design within existing practices of the users~\cite{knowles2014patterns,strengers2011designing,hansson2021decade}. We build on this work by incorporating previously used eco-feedback design elements within existing LLM user practices.  

\subsection{Environmental Psychology and Personal Norms}

Eco-feedback systems are often designed using techniques from HCI and environmental psychology~\cite{Froehlich2010_DesignEcoFeedback}. One of the most direct and powerful psychological drivers of pro-environmental behavior is \textit{personal norms}, which represent an individual's internal moral obligation~\cite{ai2024norm,de2021listen}.  
Accordingly, researchers have increasingly incorporated personal norms into eco-feedback systems, utilizing information gathered from user surveys or system telemetry~\cite{Xu2021UrbanEcoFeedback}. For example, prior works include design of eco-feedback systems that utilize a driver's personal braking style to minimize the battery consumption of EV driving ~\cite{di2017vehicle}, provide higher game rewards for users who do not frequently use a food waste digester when interacting with the system~\cite{tsita2025ai}, and determine if users have ``weak'' versus ``strong'' energy-saving attitudes based on a survey of their existing energy-saving practices~\cite{starke2020little}. Our works build on this prior research by investigating how personal norms may affect the adoption of eco-friendly LLM behavior through responses from an in-system user survey.

\subsection{Sustainable and Carbon-Aware Computing: Architecture, Systems, Design}

Researchers are increasingly exploring efforts to create more sustainable infrastructure and processes to support increasing LLM usage, particularly within service-level agreements (SLA). For example, some researchers have focused on shifting data centers to renewable energy sources and utilizing intermittent renewable energy to ensure continuous renewable energy power for data centers~\cite{acun2023carbon,wu2022sustainable}. Additional research focuses on designing energy-efficient LLM clusters via techniques such as phase splitting~\cite{patel2024splitwise}. We extend the area of carbon-aware computing for LLMs by focusing on response time latencies within SLA constraints.

\section{LLM Eco-Feedback Interface}
\label{sec:Interface}
In this section, we present the eco-feedback LLM interface. We first describe the design of ``eco modes'' that increase latency to reduce carbon emissions (Section \ref{sec:eco-mode-design}). We then describe the user interface and specific design elements to inform users of carbon outputs (Section \ref{sec:interface}).

\subsection{Eco Modes Design} 
\label{sec:eco-mode-design}

We designed the eco modes of our LLM interface around latency-based trade-offs within Service Level Agreement (SLA) constraints, allowing users to reduce carbon emissions by accepting controlled delays in response generation.

\subsubsection{LLMs and Service Level Agreements}
LLM-based services typically operate under Service Level Agreements (SLAs) that define performance expectations such as latency and availability \cite{wikiSLA}. Among these, latency is a key determinant of perceived usability. In LLM systems, latency can be characterized by time to first token (TTFT), which reflects initial responsiveness; token-by-token latency (TBT), which indicates the latency between tokens; and end-to-end latency, which captures total request-to-response time \cite{Hathora2023, AIMultiple2024}. Optimizing for sustainability involves balancing these latency components against performance expectations, motivating our design of eco modes that explicitly expose this trade-off.


\subsubsection{Defining Eco Modes}
Our interface manages response latency within SLA constraints to balance user experience and carbon emissions. Users can choose among multiple eco modes, each characterized by a distinct latency profile and corresponding carbon impact based on backend hardware configurations. To simulate latency–emissions trade-offs, we adopt three techniques from prior work that model emission reductions through controlled latency degradation~\cite{Kim2025slower}:
\begin{enumerate}
    \item \textit{Tensor Parallelism}: 
    We reduce the number of GPUs the LLM is deployed on, thus utilizing fewer hardware resources. This results in reduced carbon emissions and longer processing times. Tensor parallelism influences both TTFT and TBT~\cite{dynamollm}.
    \item \textit{Batching}: We batch incoming requests, which increases the GPU processing time but reduces the carbon emission for each request.  Therefore, the user experiences increased latency but reduces the carbon emission on the request they batched with other requests. Batching influences both TTFT and TBT.
    \item \textit{Renewable Energy Scheduling (RES)}: We send the request to a remote data center that is powered by renewable energy rather than a nearby data center powered by carbon-intensive electricity from the grid. 
    We estimate that RES adds a 150ms delay to the TTFT and decreases the carbon emissions from 380 to 295 gCO2e per kWh if the package is routed to the EU from the US~\cite{act}. 
\end{enumerate}

Because the influence of the above techniques can vary greatly with different input and output token sizes, we simulate the LLM (google/gemma-2-27b-it) across a wide range of possible input and output tokens on 1, 2, 4, and 8 GPUs with batch sizes of 1 up to 256 to get the corresponding TTFT and TBT time. We assume the default response returned from the serverless API is fixed to a GPU size of 4 and batch size of 1 with no renewable energy scheduling. We then design five eco mode configurations with increasing latency degradation and descending embodied carbon emission, as summarized in Table~\ref{tab:eco_mode_summary}.

\begin{table}[t]
\caption{Summary of eco modes and latency trade-offs. TTFT and TBT are shown as ratios normalized to the default mode.}
\centering
\renewcommand{\arraystretch}{1.1}
\begin{tabular}{lccccc}
\toprule
\textbf{Mode} & \textbf{GPUs} & \textbf{Batch} & \textbf{TTFT} & \textbf{TBT} & \textbf{RES} \\
\midrule
\textbf{Default} & 4 & 1   & 1.00 & 1.00 & $\times$ \\
1                & 4 & 2   & 1.00 & 1.89 & $\checkmark$ \\
2                & 2 & 8   & 1.83 & 3.75 & $\checkmark$ \\
3                & 2 & 32  & 2.07 & 4.80 & $\checkmark$ \\
4                & 2 & 64  & 2.37 & 5.99 & $\checkmark$ \\
5                & 2 & 128 & 2.91 & 10.32 & $\checkmark$ \\
\bottomrule
\label{tab:eco_mode_summary}
\end{tabular}
\end{table}


\subsubsection{Estimating Latencies}
Based on the simulation of input and output token sizes, as well as the number of GPUs and batch sizes used, we construct the eco mode configurations and summarize their corresponding TTFT and TBT ratios in Table~\ref{tab:eco_mode_summary}.


\begin{equation}
\begin{aligned}
t_{\text{TTFT}} = & \: t_{\text{TTFT default}} \times \text{TTFT Ratio} \\
& + \begin{cases}
150 \text{ ms}, & \text{if RES is used} \\
0, & \text{otherwise}
\end{cases}
\end{aligned}
\end{equation}


\begin{equation}
    t_{TBT} = t_{TBT default} \times \text{TBT Ratio}
\end{equation}

\begin{equation}
    t_{ETE}
= t_{TTFT} + t_{TBT} \cdot T_{\text{out}}
\end{equation}

The query runtime is characterized by the end-to-end latency $t_{ETE}$, shown in (3). This value is composed of the time-to-first-token $t_{\mathrm{TTFT}}$ (1), the per-token latency $t_{\mathrm{TBT}}$ (2) multiplied by the number of output tokens $T_{\text{out}}$, and, if applicable, an additional $150$ ms overhead when renewable energy scheduling (RES) is used.

\subsubsection{Estimating Carbon Emissions}
Based on the eco mode configurations, we calculate the carbon emission of a query using the following equations:

\begin{equation}
\begin{split}
    \text{CE}_{\text{emb}} = & \left( \frac{t_{\text{TTFT}} + t_{\text{TBT}} \times T_{\text{out}}}{LT \times B} \right) \\
    & \times (N_{\text{CPU}} \text{CE}_{\text{embCPU}} + N_{\text{GPU}} \text{CE}_{\text{embGPU}})
\end{split}
\end{equation}

\begin{equation}
\begin{split}
    \text{CE}_{\text{op}} = & \left( \frac{t_{\text{TTFT}} + t_{\text{TBT}} \times T_{\text{out}}}{B} \right) \times \text{CI}_{\text{loc}} \\
    & \times ( \text{TDP}_{\text{G}} \text{Util}_{\text{G}} N_{\text{G}} + \text{TDP}_{\text{C}} \text{Util}_{\text{C}} N_{\text{C}} )
\end{split}
\end{equation}

\begin{equation}
    \text{CE}_{\text{total}} = \underbrace{\text{CE}_{\text{emb}}}_{\text{embodied}} + \underbrace{\text{CE}_{\text{op}}}_{\text{operational}}
\end{equation}




$\text{CE}_{\text{total}}$, shown in (6) denotes the total carbon emission per query, decomposed into two components: $\text{CE}_{\text{emb}}$ (the embodied carbon emission allocated to a query, representing the amortized impact of hardware manufacturing and deployment, shown in (4)) and $\text{CE}_{\text{op}}$ (the operational carbon emission per query, representing the emissions from electricity consumption during inference, shown in (5)).  The variables $B$ and $LT$ denote the batch size (number of queries processed simultaneously) and the device lifetime, typically 5 years in data center~\cite{act}, used for amortizing embodied emissions. Device counts are represented by $N_{\text{CPU}}$ and $N_{\text{GPU}}$, while $\text{TDP}_{\text{CPU}}$ and $\text{TDP}_{\text{GPU}}$ capture the thermal design power of CPUs and GPUs, respectively, scaled by their average utilization factors ($\text{Util}_{\text{CPU}}$, $\text{Util}_{\text{GPU}}$) to reflect effective energy draw. 



\begin{figure}
    \centering
    \includegraphics[width=\columnwidth]{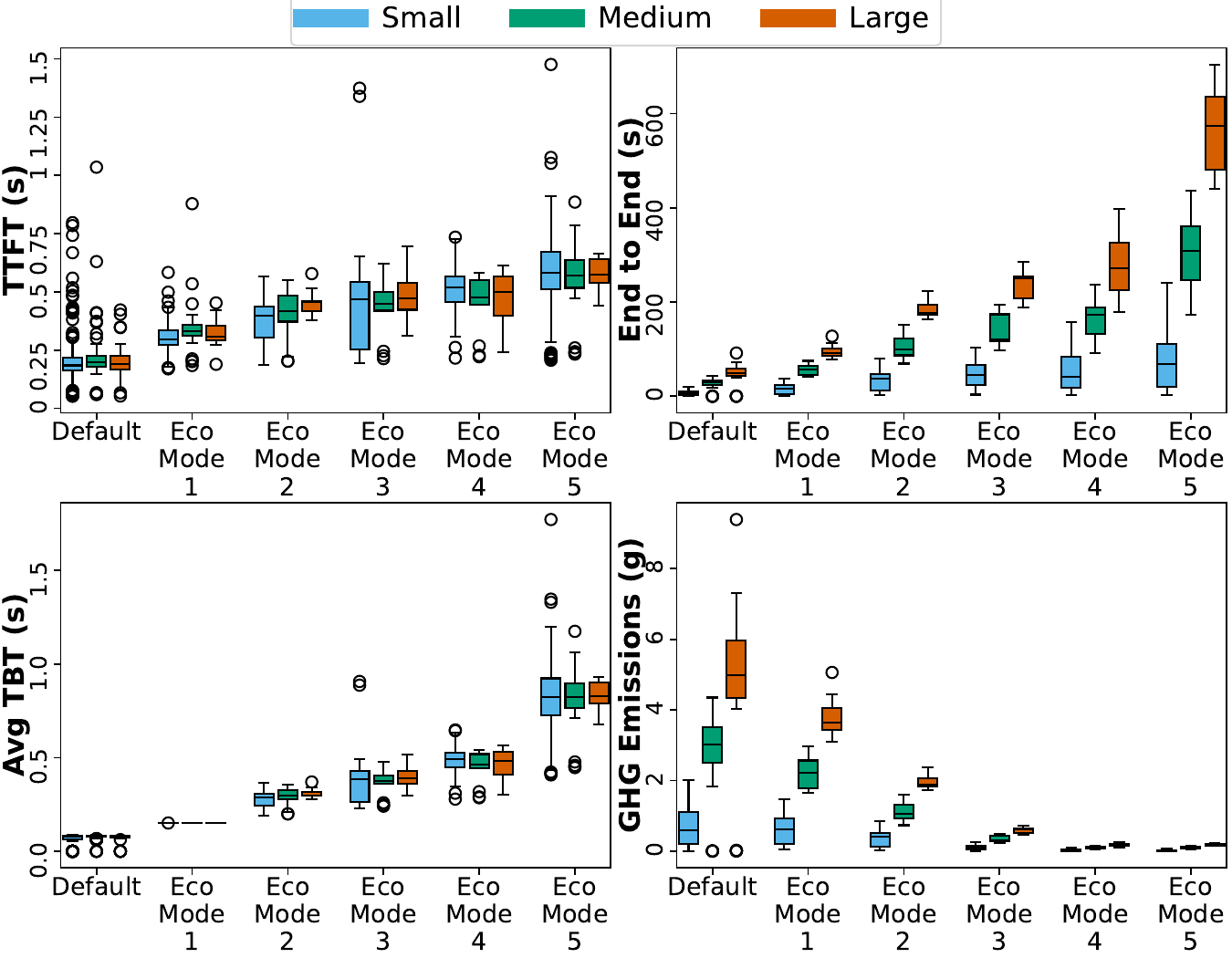}
   \caption{Latency and GHG emissions across eco modes, grouped by output token length. As eco mode increases, latency rises while emissions decrease monotonically. Larger requests (orange) show higher latency and variability than medium (green) and small (blue), highlighting the latency--emissions trade-off at higher eco modes.}
    \label{fig:latency+ghg_vs_ecomode}
\end{figure}

\subsubsection{Latency and GHG Emission Evaluation across Eco Mode}
To illustrate the latency and GHG emissions across different eco modes, we plot TTFT, average TBT, end-to-end latency, and GHG emissions in Fig.~\ref{fig:latency+ghg_vs_ecomode}. However, both end-to-end latency and GHG emissions are strongly influenced by the input and output token lengths—the longer the tokens, the higher the latency and emissions. Therefore, following a similar practice to DynamoLLM~\cite{dynamollm}, we categorize requests into three groups based on output token length: \emph{small} ($<256$ tokens), \emph{medium} ($256 \leq \mathrm{length} < 512$ tokens), and \emph{large} ($\geq512$ tokens).
We base this categorization solely on output token length, as 98.2\% of participant-provided inputs contain fewer than 128 tokens, making input lengths relatively small and consistent. Results in Fig.~\ref{fig:latency+ghg_vs_ecomode} show consistent TTFT and average TBT increases with Eco Mode number, as designed in Table~\ref{tab:eco_mode_summary}. 
To make these trade-offs more explicit, Table~\ref{tab:eco_mode_summary} summarizes the design-level configuration and normalized TTFT/TBT trade-offs for each eco mode, while Fig.~\ref{fig:latency+ghg_vs_ecomode} shows the observed latency and GHG emission distributions across request sizes. For \emph{small} requests, latency remains lowest with a tight spread; \emph{medium} requests shift upward; and \emph{large} requests are consistently the slowest and most variable, with differences widening at higher eco Modes, especially for end-to-end latency. It is worth noting that while TTFT and average TBT latency show variability within the seconds range, end-to-end latency can vary much more widely, reaching up to 252 seconds, especially for large requests. This is largely driven by variation in output token length, which depends on the specific questions users ask. Participants in the live system directly experienced these delays, and the exact end-to-end latency varied across interactions based on the submitted prompts and resulting output lengths. GHG emissions decrease monotonically as eco Mode increases for all sizes, yet remain highest for large and lowest for small; the largest absolute reduction occurs for large requests, highlighting a clear latency and emissions tradeoff across request sizes.


\begin{figure*}
    \centering
    \includegraphics[width=\linewidth]{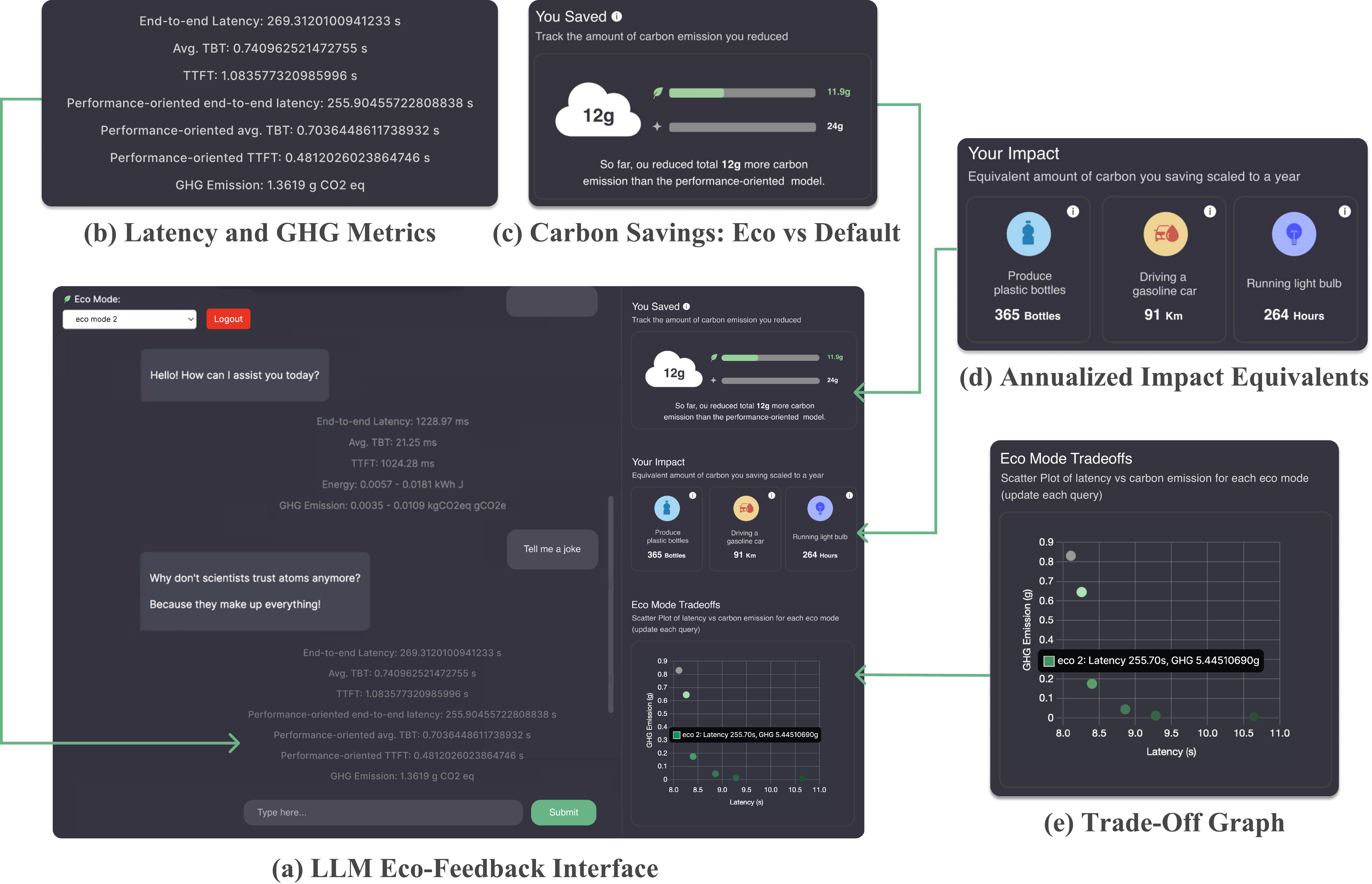}
    \caption{Representative LLM eco-feedback interface and visualization components used in the experiment.
(a) Experimental interface with default and eco-mode selection via a dropdown menu.
(b) Response-level latency and GHG metrics, showing real-time latency and GHG feedback with cumulative savings.
(c) \textit{You Saved} (carbon reduction vs.\ default).
(d) \textit{Your Impact} (annualized environmental impact shown through real-world equivalence formats) \cite{dean2024chatgpt, Franklin2011, SavingLightBulb2018, dieselnet_ghg_standards}.
(e) \textit{Trade-off Graph} showing a scatterplot of latency and carbon emissions across modes.
Hovering over exclamation icons provides further explanations.}

    \label{fig: testing setting}
    \end{figure*}

\subsection{Interface Design}
\label{sec:interface}


We designed an eco-feedback interface to reduce LLM-related carbon emissions while engaging users in sustainable decision-making. Guided by the question: How might we design an eco-feedback interface for LLM systems that raises awareness and promotes sustainable behavior through flexible, low-emission usage options ~\cite{InteractionDesignHMW}, we reviewed prior work on eco-feedback and data visualization strategies ~\cite{Mahyar2024, Chalal2022} 
and iteratively developed a Figma prototype supported by a serverless API. Participants interacted with a live LLM system that produced real eco-mode latency differences. The interface adopts a familiar LLM chat layout to minimize learning overhead while introducing multiple eco modes that trade response latency for lower emissions.

As shown in Fig.~\ref{fig: testing setting}, users choose between a default mode and five eco modes, each providing real-time feedback on latency and estimated greenhouse gas emissions. Four complementary eco-feedback visualizations—detailed metrics, relative savings, real-world impact equivalents, and a latency–emissions trade-off graph—support both immediate awareness and comparison of sustainability trade-offs. 

Your Impact (Fig.~\ref{fig: testing setting}d) contextualizes annualized savings, assuming 1,217 queries per year based on a conservative estimate of daily LLM use reported by recent traffic statistics~\cite{dean2024chatgpt}. We report annualized values because per-query savings are very small and may be difficult to interpret meaningfully. The interface presents these savings through three familiar equivalence formats—PET bottle production, gasoline car travel, and light bulb usage—to make environmental impact information more relatable to users~\cite{Franklin2011,SavingLightBulb2018,dieselnet_ghg_standards}.

\section{Methodology}
\label{sec:methodology}


\subsection{Participants}
We recruited 102 undergraduate students enrolled in a mandatory Computer Ethics course in a university Computer Science program.
After excluding incomplete or invalid responses, we retained data from 89 participants (ages 18–24, U.S.-based) for analysis. All participants provided informed consent, and the study protocol was approved by the institution’s Institutional Review Board (IRB). We focused on Computer Science undergraduates as technologically literate end-users with frequent exposure to LLMs who also represent future AI developers, a choice consistent with prior work showing high levels of generative AI adoption among young adults~\cite{Young2024role}. This demographic is also well suited for examining sustainability-driven interfaces, as prior research suggests that young adults (18–24) exhibit relatively strong environmental concern and engagement~\cite{Latkin2021correlates, romano2024, pew2021}.

\subsection{Experimental Procedure} 
We conducted the user study over three days across ten sections of a university Computer Science ethics course. Participants used their own laptops, and course teaching assistants supported the study. Participants logged into the MVP eco-feedback interface using individual credentials. Participants interacted with a functional LLM system during the study. Within the interface, system configurations were presented via a dropdown menu labeled as Eco Mode 1–5, making the sustainability framing visible to participants. However, the specific implications of each mode, including the magnitude of latency and emission trade-off, were not explained in advance and were instead revealed through direct interaction and feedback during use. 

\subsubsection{Individual Survey}
Participants first explored the interface freely and then completed a set of randomized interaction tasks, followed by an online survey. The survey was designed to assess participants' understanding of the eco-feedback interface, their experience, and the effectiveness of eco-feedback data in communicating environmental impact. It combined Likert-scale questions, open-ended responses, and ranking tasks to capture quantitative and qualitative insights.

For each of the five eco modes, participants indicated their preference between the default and eco mode, rated satisfaction, and reported how perceived latency, carbon savings, and performance differences influenced their choices. Participants also evaluated four eco-feedback visualizations (Fig.~\ref{fig: testing setting}) by reporting attention, satisfaction, and perceived impact. The survey further measured the likelihood of future use and personal environmental norms using a validated 10-item Likert-scale instrument. A summary of the survey materials is provided in Appendix~\ref{appendix:user_study_materials}.


\subsubsection{Group Discussion}

Following the survey, we conducted a group discussion to elicit qualitative reflections on usability, interpretation of eco-feedback, and design suggestions. These discussions were audio-recorded with consent and complemented the survey by capturing motivations and interpretations not evident in structured responses.

\subsection{Data Analysis} 
We analyzed the data using both qualitative and quantitative methods. For open-ended survey responses and discussion data, we conducted a reflexive thematic analysis following Braun and Clarke’s approach~\cite{Braun2019ReflexiveTA}. The first author performed initial coding and codebook development, which was iteratively refined and reviewed by three additional researchers. Disagreements were resolved through discussion. This process produced 85 initial codes that were consolidated into 36 sub-themes and four higher-level themes describing user perceptions of the eco-feedback interface.

For quantitative survey data, we used R to examine relationships between eco-mode preference and perceived latency, carbon-saving impact, default-mode differences, and environmental norms. Likert-scale responses were treated as ordered indicators rather than interval measures, and inferential claims were limited accordingly; satisfaction and ranking items are reported descriptively. We calculated the following statistics:

\begin{itemize}

   \item \textbf{Generalized Linear Mixed-Effects Model (GLMM):} We modeled eco-mode preference (eco vs.\ default) using a GLMM with a logit link and participant-level random intercepts. Fixed effects included eco mode, perceived carbon-saving impact, latency, perceived performance differences, and personal norms. We report results as odds ratios with 95\% confidence intervals and $p$-values.

    \item \textbf{Descriptive Statistics:} Frequencies, percentages, means, and standard deviations summarize adoption intent, satisfaction, and visualization rankings.
\end{itemize}

\section{Findings}
\label{Results}
In this section, we present results organized around our research questions, applying the methodology described in Section~\ref{sec:methodology} to the system outlined in Section~\ref{sec:Interface}.




\subsection{User Preference and Satisfaction of Eco Modes}

\begin{figure}[t]
    \centering
    \includegraphics[width=\columnwidth]{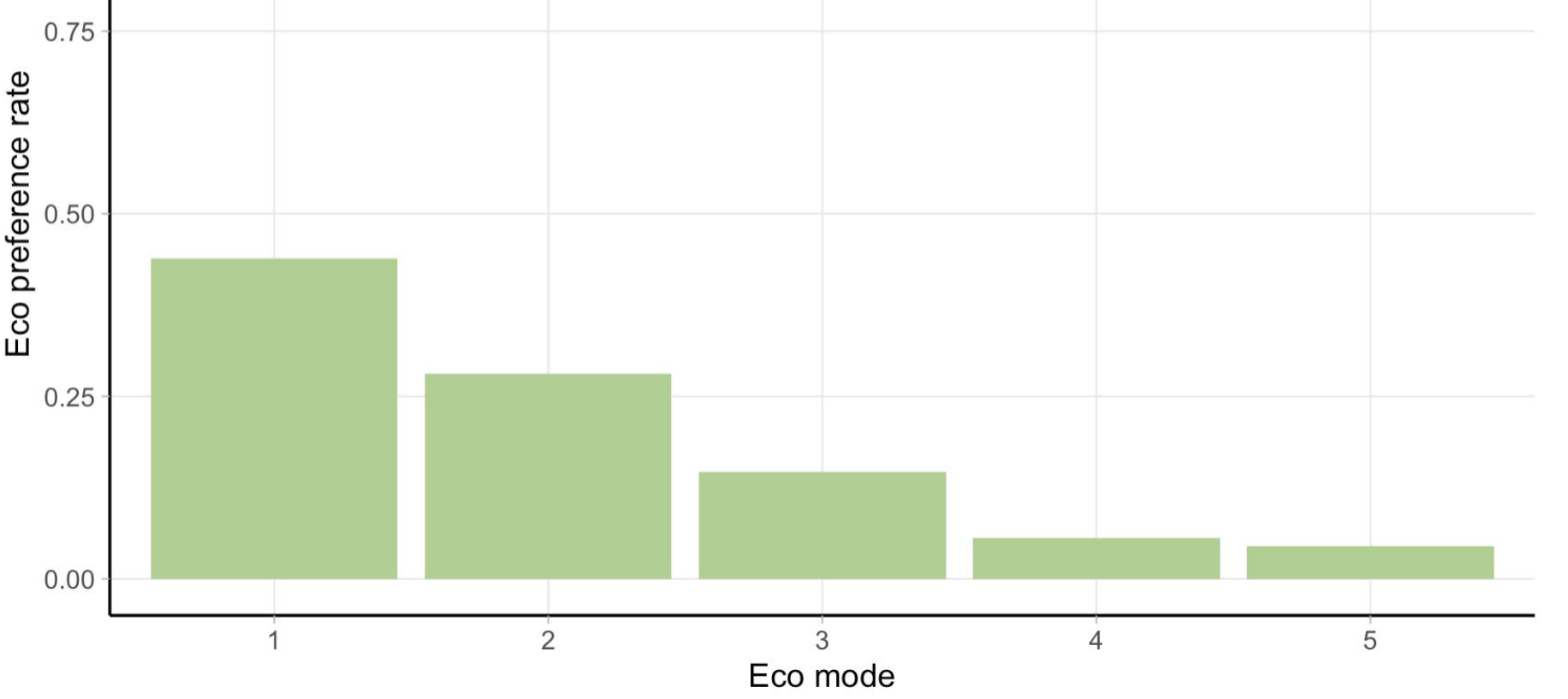}
    \caption{Proportion of interactions in which participants selected the eco mode rather than the default mode across system modes, from Mode 1 (lowest latency) to Mode 5 (highest latency).}
    \label{fig:Eco preference}
\end{figure}

As shown in Fig.~\ref{fig:Eco preference}, the proportion of interactions in which participants selected the eco mode over the default mode decreased sharply from eco mode~1 to eco mode~5, dropping from approximately 45\% in eco mode~1 to below 5\% in eco mode~5. 
This pattern is statistically supported by the mixed-effects logistic regression results in Table~\ref{tab:regression_results} A. 

To complement the preference analyses, we examined user satisfaction across modes.
Consistent with earlier preference findings, mode~1 showed the highest satisfaction (mean = 3.64, SD = 0.82), indicating a consistently positive experience. Modes~2 and~3 received moderate ratings (means = 3.03 and 2.76) with greater variability, suggesting more mixed evaluations. In contrast, satisfaction declined sharply for modes~4 and~5 (means = 2.35 and 2.06), with mode~5 showing the greatest dispersion (SD = 1.03), indicating that while a small subset of users tolerated highly constrained modes, most reported dissatisfaction, likely reflecting the increasing burden of response latency and reduced usability.


\subsubsection{Impact of Latency, Carbon-Saving, and performance difference}

As seen in Table~\ref{tab:regression_results} B, higher perceived latency was strongly associated with reduced eco mode preference (OR = 0.192, $p < .001$). In contrast, recognition of carbon-saving impacts emerged as the strongest positive predictor of eco mode selection, increasing the likelihood of eco preference both when modeled alone (OR = 3.879, $p < .001$), as seen in Table~\ref{tab:regression_results} C. 

Importantly, even in the combined model shown in Table~\ref{tab:regression_results} E, which accounts for users’ recognition of carbon savings, latency, and performance differences, the highest-latency configurations (modes~4–5) were still significantly less likely to be selected than mode~1 (mode~4: OR = 0.136, $p = .010$; mode~5: OR = 0.131, $p = .028$). 
It suggests that for moderately delayed modes (modes~2–3), lower eco-mode preference is largely explained by what users perceive, such as increased delay or reduced performance, rather than by the mode itself. In contrast, the delays in modes~4–5 are sufficiently large that eco-mode adoption declines regardless of users’ perceptions.
Overall, this pattern indicates that perceived environmental benefits support eco-mode adoption, but increasing response latency limits users’ willingness to engage with these options.


\begin{table}[t]
\caption{Mixed-effects logistic regression of system modes and user-recognized carbon savings, latency, and performance differences predicting preference; ORs with 95\% CIs reported.}
\centering
\label{tab:regression_results}
\scriptsize
\renewcommand{\arraystretch}{1.3}
\begin{tabularx}{\columnwidth}{@{} >{\raggedright\arraybackslash}X l l l @{}}
\toprule
\textbf{Predictor} & \textbf{OR} & \textbf{95\% CI} & \textbf{p} \\
\midrule
\multicolumn{4}{l}{\textit{A) Mode only (reference = Mode~1)}} \\
Mode 2 & 0.311 & [0.135, 0.719] & 0.006 \\
Mode 3 & 0.084 & [0.030, 0.234] & <.001 \\
Mode 4 & 0.019 & [0.005, 0.075] & <.001 \\
Mode 5 & 0.014 & [0.003, 0.061] & <.001 \\
\midrule

\multicolumn{4}{l}{\textit{B) Mode + latency recognition}} \\
Latency recognition & 0.192 & [0.106, 0.349] & <.001 \\
Mode 2 & 0.779 & [0.313, 1.939] & 0.591 \\
Mode 3 & 0.314 & [0.107, 0.926] & 0.036 \\
Mode 4 & 0.106 & [0.024, 0.462] & 0.003 \\
Mode 5 & 0.137 & [0.026, 0.722] & 0.019 \\
\midrule

\multicolumn{4}{l}{\textit{C) Mode + carbon saving recognition}} \\
Carbon recognition & 3.879 & [2.198, 6.847] & <.001 \\
Mode 2 & 0.262 & [0.105, 0.653] & 0.004 \\
Mode 3 & 0.055 & [0.017, 0.175] & <.001 \\
Mode 4 & 0.013 & [0.003, 0.061] & <.001 \\
Mode 5 & 0.006 & [0.001, 0.035] & <.001 \\
\midrule

\multicolumn{4}{l}{\textit{D) Mode + performance difference recognition}} \\
Difference recognition & 0.403 & [0.235, 0.693] & 0.001 \\
Mode 2 & 0.458 & [0.193, 1.087] & 0.077 \\
Mode 3 & 0.185 & [0.064, 0.539] & 0.002 \\
Mode 4 & 0.045 & [0.011, 0.193] & <.001 \\
Mode 5 & 0.051 & [0.010, 0.260] & <.001 \\
\midrule

\multicolumn{4}{l}{\textit{E) Combined model}} \\
Carbon recognition & 5.016 & [2.774, 9.067] & <.001 \\
Latency recognition & 0.198 & [0.103, 0.378] & <.001 \\
Difference recognition & 0.544 & [0.292, 1.015] & 0.056 \\
Mode 4 & 0.136 & [0.030, 0.625] & 0.010 \\
Mode 5 & 0.131 & [0.021, 0.803] & 0.028 \\

\bottomrule
\end{tabularx}

\vspace{0.5em}
\footnotesize{
Models include participant-level random intercepts to account for repeated measures. In the combined model, only system modes with statistically significant effects are reported after accounting for users’ recognition of carbon savings, latency, and performance differences.
}
\end{table}

\subsubsection{Influence of Eco-mindedness}

\begin{table}[t]
\caption{GLMM predicting eco-mode preference (eco = 1) from system modes and personal norm groups; ORs with 95\% CIs reported.}
\centering
\label{tab:glmm_or_ci_p_clean}
\scriptsize
\setlength{\tabcolsep}{3pt}
\renewcommand{\arraystretch}{1.15}

\begin{tabularx}{\columnwidth}{@{}
  >{\raggedright\arraybackslash}X
  >{\centering\arraybackslash}p{0.40\columnwidth}
  >{\raggedleft\arraybackslash}p{0.12\columnwidth}
@{}}
\toprule
\textbf{Predictor} & \textbf{OR [95\% CI]} & \textbf{p} \\
\midrule
Intercept (Mode~1, Low norm) & 0.075 [0.010, 0.562] & 0.012 \\

Mode~2 (vs.~Mode~1) & 0.500 [0.048, 5.186] & 0.562 \\
Mode~3 & 0.178 [0.011, 2.964] & 0.229 \\
Mode~4 & 0.000 [0, 6.23e81] & 0.860 \\
Mode~5 & 0.000 [0, 4.24e78] & 0.855 \\

Medium norm (31--38) & 8.951 [1.040, 77.061] & 0.046 \\
High norm (39--47) & 50.339 [3.013, 840.984] & 0.006 \\

Mode~2 $\times$ Medium norm & 0.493 [0.039, 6.313] & 0.587 \\
Mode~3 $\times$ Medium norm & 0.516 [0.025, 10.552] & 0.668 \\
Mode~4 $\times$ Medium norm & 3.33e6 [0, 2.69e96] & 0.887 \\
Mode~5 $\times$ Medium norm & 2.21e6 [0, 1.21e93] & 0.886 \\

Mode~2 $\times$ High norm & 0.728 [0.034, 15.697] & 0.839 \\
Mode~3 $\times$ High norm & 0.130 [0.003, 5.269] & 0.280 \\
Mode~4 $\times$ High norm & 3.74e5 [0, 3.09e95] & 0.903 \\
Mode~5 $\times$ High norm & 3.72e5 [0, 2.05e92] & 0.900 \\
\bottomrule
\end{tabularx}

\vspace{0.5em}
\footnotesize{Personal norm scores  (max = 50) were grouped by interquartile range into Low (21–30), Medium (31–38), and High (39–47); models include participant random intercepts; large ORs reflect sparse eco-mode selections.}
\end{table}
To understand how pro-environmental orientations shape engagement with the interface, we surveyed participants about their personal environmental practices then compared eco mode preferences across ``personal-norm'' groups mapped to their eco-mindedness using a mixed-effects model (Table~\ref{tab:glmm_or_ci_p_clean}). 
In eco mode~1, participants with medium and high levels of eco-mindedness were significantly more likely to select eco modes (Medium: OR = 8.95, $p = .046$; High: OR = 50.34, $p = .006$) than those with low eco-mindedness. Yet higher-latency modes showed very low baseline selection odds relative to mode~1 (mode~4–5 OR $\approx$ 0, $p > .85$), reflecting low preference for those eco modes. These results indicate that stronger environmental norms raise initial eco mode willingness, but this advantage diminishes as response delays grow.

\subsubsection{Likelihood of Long-Term Adoption}

To assess the likelihood of adoption of an eco-feedback LLM, participants rated their willingness to use such a system on a 5-point Likert scale. Among the 89 participants, 5.6\% reported being very likely to use the system, 16.9\% somewhat likely, and 12.4\% neither likely nor unlikely. In contrast, 48.3\% indicated being somewhat unlikely and 16.9\% extremely unlikely to adopt the system for regular use. These responses show that a majority of participants expressed reluctance toward adoption, with nearly two-thirds reporting some degree of unwillingness.

\subsection{Effectiveness of Eco-Feedback Visualizations}


\begin{figure}[t]
    \centering
    \includegraphics[width=\columnwidth]{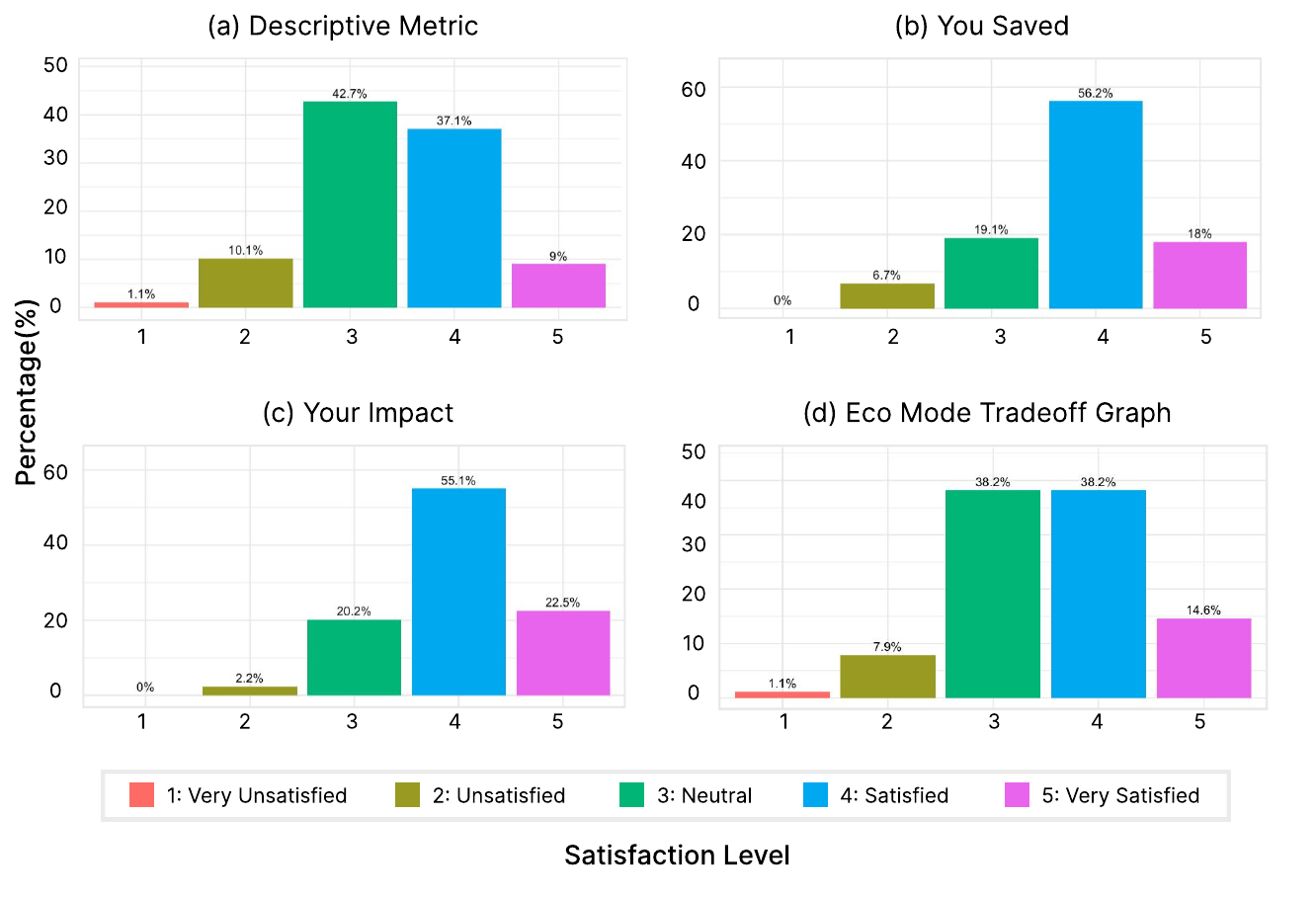}
    \vspace{-0.8cm}
    \caption{Bar charts show user satisfaction distributions for four real-time eco-feedback designs on a 5-point Likert scale.}
    \label{fig:data satisfaction}
\end{figure}

To assess how eco-feedback visualizations shape user understanding and engagement, we analyzed participants’ attention, satisfaction, and preferences across the four visualization formats shown in the interface (Fig.~\ref{fig: testing setting} b-e), using Likert-scale responses and qualitative feedback. Satisfaction and ranking measures are reported descriptively to summarize trends rather than to support inferential comparisons. Most participants noticed the eco-feedback elements, with 80.9\% reporting that they paid at least some attention to the displayed information.

As shown in Fig.~\ref{fig:data satisfaction}, participants expressed generally high satisfaction across all four formats. In particular, the more interpretive and outcome-oriented visualizations were better received than purely descriptive metrics. Within the Your Impact visualization, real-world analogies framing carbon savings as everyday activities (e.g., driving a gasoline car or running a light bulb) were rated more favorably than abstract units, indicating that familiar environmental comparisons helped participants make sense of sustainability information. These findings suggest that providing eco-feedback information can support user engagement and is most effective when environmental benefits are presented clearly, concretely, and with minimal technical complexity. 

\subsection{User Responses and Learning Effects}

\renewcommand{\arraystretch}{1.12}
\setlength{\tabcolsep}{5.4pt}
\setlength{\arrayrulewidth}{0.1pt}

\begin{table*}[t]
\centering
\footnotesize
\caption{Qualitative themes in eco-mode adoption: Facilitators, barriers, and educational impact.}
\label{tab:user_perceptions_educational_integrated}

\begin{tabularx}{\textwidth}{|>{\raggedright\arraybackslash}p{1.70cm}
|>{\raggedright\arraybackslash}p{2.50cm}
|>{\raggedright\arraybackslash}p{5.95cm}
|>{\raggedright\arraybackslash}X|}

\noalign{\global\arrayrulewidth=0.6pt}\hline
\rowcolor{gray!16}
\textbf{Analytic Role} & \textbf{Theme} & \textbf{Interpretation} & \textbf{Representative Quotes} \\
\noalign{\global\arrayrulewidth=0.6pt}\hline
\noalign{\global\arrayrulewidth=0.3pt}

\multirow[t]{3}{=}{\textbf{Motivating Eco-Mode Choice}}
& Environmental Awareness
& Participants recognized the sustainability-oriented purpose of the interface and viewed environmental feedback as relevant to interaction.
& ``It felt like it was tracking my environmental impact while I was using it.''  
``I liked that it showed something related to carbon usage or saving.'' \\
\cline{2-4}

& Eco-Conscious Framing of AI Use
& The interface reframed everyday LLM usage as environmentally consequential, encouraging sustainability reflection.
& ``My first thought was that it was going to be about saving energy.''  
``It made me think about the environmental impact of using AI.'' \\
\cline{2-4}

& Relatable Eco-Feedback
& Familiar representations helped users interpret abstract carbon metrics and perceive their actions as meaningful.
& ``Show me what my usage means in a real-world way.''  
``It feels good to know I’m not being as wasteful.'' \\
\noalign{\global\arrayrulewidth=0.35pt}\hline
\noalign{\global\arrayrulewidth=0.1pt}

\multirow[t]{2}{=}{\textbf{Shaping User Expectations}}
& Familiarity with Existing LLMs
& Participants compared the system to familiar tools (e.g., ChatGPT), shaping expectations about speed and usability.
& ``It looks like ChatGPT but with eco features.''  
``I expected it to behave like other chatbots.'' \\
\cline{2-4}

& Interface Design and Aesthetics
& The interface was perceived as clean and simple, though some felt it lacked clear cues differentiating eco modes.
& ``The interface looked very simple and easy to use.''  
``It was clean, but I wasn’t sure what made it different.'' \\
\noalign{\global\arrayrulewidth=0.35pt}\hline
\noalign{\global\arrayrulewidth=0.1pt}

\multirow[t]{4}{=}{\textbf{Constraining Eco-Mode Effectiveness}}
& Metric Confusion and Jargon
& Technical terminology and unclear indicators reduced interpretability.
& ``I don’t know what most of those acronyms mean.''  
``I didn’t really understand what the numbers meant.'' \\
\cline{2-4}

& Confusion About Eco-Modes
& Participants struggled to intuitively understand differences between modes beyond response speed.
& ``I wasn’t sure what each eco mode did.''  
``I didn’t know what changed between modes.'' \\
\cline{2-4}

& Response Latency
& Delayed responses disrupted engagement and outweighed sustainability motivation in higher modes.
& ``It was way too slow.''  
``Eco mode 1 was the only one usable.'' \\
\cline{2-4}

& Limited User Control
& Lack of interaction control amplified frustration under delay.
& ``I wish there was a stop button.''  
``It felt hard to stay focused while waiting.'' \\
\noalign{\global\arrayrulewidth=0.35pt}\hline
\noalign{\global\arrayrulewidth=0.1pt}

\multirow[t]{2}{=}{\textbf{Educational Potential}}
& Bridging Technical Literacy Gaps
& Analysis revealed a significant baseline gap in understanding GHG and latency as environmental factors, which the interface began to fill.
& ``I never thought about latency as an environmental factor; I just thought it was a slow connection.''
``I don't know what GHG stands for in this context.'' \\
\cline{2-4}

& Understanding Environmental Impact
& The interface served as a scaffolding tool, helping users learn that digital actions have physical resource implications.
& ``I learned that even a simple chat has a carbon emission.''
``It helped me understand the trade-off between being fast and being green.'' \\
\noalign{\global\arrayrulewidth=0.35pt}\hline
\noalign{\global\arrayrulewidth=0.1pt}

\multirow[t]{2}{=}{\textbf{Opportunities for Improvement}}
& Desire for Actionable Eco-Feedback
& Participants wanted guidance on how to act on eco-feedback beyond awareness.
& ``It would be nice if it gave tips or suggestions to save more carbon.'' \\
\cline{2-4}

& Personalization and Real-World Context
& Users suggested more personalized or real-world analogies to improve understanding and credibility.
& ``Like comparing it to water or energy saved in daily life.''  
``Show me what my usage means in real-world terms.'' \\
\noalign{\global\arrayrulewidth=0.35pt}\hline
\end{tabularx}
\end{table*}

We analyzed open-ended responses 
to examine how participants understood and responded to the eco feedback interface. Table~\ref{tab:user_perceptions_educational_integrated} summarizes qualitative themes that explain eco mode use and reveal learning related effects observed during interaction. 
It shows that participants interpreted the eco-feedback interface as both an interaction aid and a source of new understanding about environmental impact. Several participants immediately recognized the sustainability focus of the system, describing it as “tracking my environmental impact” and noting that it “made me think about the environmental impact of using AI.” Relatable feedback further helped participants connect abstract metrics to meaningful outcomes, with one participant stating, “It feels good to know I’m not being as wasteful.”

At the same time, responses revealed gaps in participants' prior understanding of digital sustainability. Technical terms and metrics were frequently described as confusing, with participants stating, “I don’t know what most of those acronyms mean” and “I never thought about latency as an environmental factor; I just thought it was a slow connection.” Several participants explicitly reported learning through the interface, noting that “even a simple chat has a carbon emission” and that it helped them understand “the trade off between being fast and being green.”

Participants also interpreted the system through comparisons with existing language model tools. Descriptions such as “it looks like ChatGPT but with eco features” shaped expectations around speed and usability, and in some cases contributed to uncertainty about differences between eco modes beyond response delay. As interaction continued, response latency emerged as a dominant constraint. Higher eco modes were often described as unusable, with participants stating “Eco mode 1 was the only one usable” and “it was way too slow,” particularly when combined with limited interaction control, such as the inability to stop responses while waiting. 

These findings suggest that interacting with the eco-feedback interface exposed gaps in participants’ understanding of digital sustainability and supported learning about how everyday LLM use relates to environmental impact, within clear usability limits.

\section{Discussion}



In this section, we discuss the implications of our results for designing sustainable eco-feedback interfaces in LLM systems, as well as study limitations and future directions.

\subsection{Opportunities for Sustainability and Education}
\subsubsection{Sustainability Within Acceptable Carbon Saving}
Consistent with prior work~\cite{Kim2025slower}, our findings confirm that users tolerate modest latency when environmental benefits are made explicit, highlighting the promise of latency-aware eco-feedback for supporting more sustainable interaction. In our system, eco mode~1 reduced emissions by about 12\,g CO$_2$ per query, or 14.6\,kg CO$_2$ annually per user assuming 1,217 queries per year. Given its 43\% selection rate in our study, eco mode~1 suggests meaningful carbon-saving potential when lower-impact modes remain within acceptable latency bounds. For comparison, a typical passenger vehicle emits about 4.6 tons of CO$_2$ annually~\cite{BusinessInsider2025, TheTech2025, EducationData2026}.


However, perceived latency remained the primary barrier to sustained eco-mode use. Adoption depended not only on carbon-awareness cues but also on how noticeable delays felt during interaction. Interfaces may therefore improve tolerance for lower-carbon modes by making wait time more manageable or by emphasizing carbon-saving benefits more clearly. Prior work suggests that interface and system-level strategies such as progress indicators, lightweight waiting interactions (e.g., the Chrome offline dinosaur game)~\cite{GoogleDinoBlog2018, WikipediaDinosaurGame}, or partial output streaming and hybrid response architectures~\cite{Adenekan2023_OptimizingLLM} can help reduce perceived wait time.

Latency also reshaped user behavior rather than simply deterring use. Some participants shortened prompts or adjusted interaction patterns to maintain control under delay, underscoring the importance of preserving user agency through visible system status and anticipatory feedback. This aligns with HCI findings that progress signals increase tolerance for wait time~\cite{Nah2004TolerableWait}.

\subsubsection{Eco-Feedback as a Learning-Oriented Interaction}
Qualitative analysis suggests that eco-feedback can support learning by shaping how users reason about the environmental implications of LLM use. Eco-feedback helped make otherwise invisible aspects of digital sustainability more legible during routine interaction. In particular, concrete and interpretable feedback supported reflection on how everyday choices, such as prompt length or mode selection, relate to environmental impact. In this sense, eco-feedback functions not as formal instruction, but as an interactional cue that encourages sense-making during use.

From an educational perspective, this positions eco-feedback interfaces as lightweight, experiential learning touchpoints embedded in everyday AI use. By situating environmental information directly within interaction, eco-feedback supports learning that emerges through use rather than relying on explicit instruction. Consistent with prior work on practice-oriented environmental education, such situated and feedback-driven interaction can support emerging understanding of environmental issues and foster awareness of how individual actions relate to sustainability outcomes~\cite{reunila2018experience}.

\subsection{Interface Design Considerations}
\subsubsection{Eco-Feedback Interpretability}
Our results show that perceived carbon-saving effects and visible eco-feedback both supported engagement with eco modes. Visualizations that translated emissions into concrete, everyday equivalents were perceived as more effective than abstract metrics, highlighting the value of relatable, outcome-focused sustainability feedback for LLM use. However, some participants were confused by common eco-feedback symbols and carbon metrics (e.g., leaf icons or savings units). Although such imagery is widely used in prior eco-feedback systems~\cite{Froehlich2010_DesignEcoFeedback, Bao2018_EmotionalEcoFeedback}, it may not fully convey the scale and resource implications of LLM use. Designers may therefore explore more expressive or provocative sustainability visualizations, such as SavetheAI~\cite{Qiao2025ThirstyAI}.

Participants also struggled with latency and trade-off metrics when presented without sufficient context. Thus, eco-feedback should prioritize information and visuals grounded in users’ everyday experiences to ensure accessibility and intuitive understanding. For example, introducing onboarding elements that explain the function of eco modes, how carbon reduction is achieved, and what information can be accessed throughout the interface would improve transparency and user confidence.

\subsubsection{Personal Norms and Personalization}
When participants used eco mode~1, personal environmental norms played a key role, with greater eco-mindedness associated with stronger eco-mode preference. However, this effect weakened substantially in higher modes, as eco-mode adoption declined even among environmentally motivated users when latency increased and interaction frictions became more pronounced. This suggests that while environmental values can support initial engagement, performance constraints quickly override norm-driven motivation.

These results suggest that personal norms are useful for identifying users’ initial willingness to engage with lower-carbon modes, but are insufficient to sustain adoption under high-delay conditions. 
Designers may consider personalization strategies that lower the barrier to sustainable action. For example, setting low-latency eco modes as defaults for users with stronger environmental norms, offering gentle prompts when time pressure is low, or framing eco mode~1 as a recommended baseline rather than an exceptional choice. However, because we did not collect qualitative data specifically about eco-mindedness, it remains unclear how such users interpret or respond to personalization, suggesting an important direction for future work.

\subsection{System Constraints and Direction}
\subsubsection{Backend Systems Changes that Influence Design}

Beyond exploring user interface design techniques that can impact perceived or true latency, researchers can also explore backend system changes to influence latency for users. In our system, the latency configurations -- especially between default, eco Mode 1, and eco Mode 2 -- are relatively coarse-grained, producing large and easily perceivable latency gaps. This limited granularity stems both from the design of the system knobs, such as the number of GPUs and batch size, and from deployment simplifications made to ensure reproducibility. Specifically, the study assumes a fixed backend configuration consisting of four GPUs, a batch size of one, and no renewable energy scheduling, whereas real-world data centers employ far more complex scheduling, batching, and resource management strategies. These design constraints and modeling assumptions result in discrete latency shifts that do not fully capture the continuous and adaptive behavior of production systems. The implications of this are seen in the results, where there is a large drop-off in eco mode preference between eco modes 1 and 2 (as seen in Fig.~\ref{fig:Eco preference}). 

Future iterations could sweep a broader set of tunable parameters, such as prefill chunk sizes, GPU frequencies, or query scheduling policy to enable smoother latency transitions and more realistic modeling of latency–carbon trade-offs under dynamic workloads. Building and evaluating such systems could reveal more precise latency thresholds that are acceptable for different users, allowing for the development of widely accepted carbon-saving LLM systems.

\subsubsection{Uncertainty in Carbon Modeling and Limitations in Impact Captured}

As prior work highlights~\cite{carbonclarity}, carbon modeling for computing systems involves uncertainty due to opaque supply-chain data, evolving hardware lifecycles, and limited transparency from cloud providers. In our study, the assumed correspondence between batch size, GPU allocation, and renewable energy scheduling further simplifies the realities of production data centers, where LLM serving relies on complex resource management. As a result, the latency–GHG relationship presented here should be interpreted as illustrating general trends rather than precise real-world performance.

Accordingly, carbon reduction estimates should be interpreted cautiously. Variability in energy sources and system-level optimizations limits the precision of inference-time carbon accounting, and overstating exact savings risks reinforcing greenwashing~\cite{earth_org_greenwashing_2024}. Eco-feedback systems should therefore emphasize relative trade-offs and transparent communication rather than precise numerical claims.



\subsection{Research Limitations and Future Work}
\subsubsection{Sample and Demographic Scope}
Our study focused on undergraduate computer science students at a single U.S. university, a group selected for their frequent LLM use and engagement with environmental issues. Although participants were recruited from a mandatory Computer Ethics course, which reduces self-selection into the course itself, this context may still limit generalizability and introduce sampling bias, as participants may be more technically informed and environmentally attentive than broader LLM user populations. The course context may also have encouraged socially desirable responses. We did not collect demographic attributes such as gender, race, or nationality, limiting analysis of how these factors may shape eco-feedback interpretation and engagement. Future work should examine these questions across broader populations and contexts.

\subsubsection{Environmental Metrics Beyond Carbon}

While this study focuses on latency and carbon interactions, other environmental dimensions such as water consumption~\cite{water} and human health impacts~\cite{health} also represent important aspects of sustainability. Extending future work to integrate these metrics could provide a more holistic view of environmental trade-offs and engage broader audiences who may be less familiar with carbon accounting but are concerned with sustainability outcomes more broadly.

\subsubsection{Interface Design Restrictions}

Several interface limitations likely influenced user satisfaction, particularly under higher-latency conditions. Limited interaction controls, such as the absence of stop or scroll functionality, compounded the perceived cost of waiting and reduced usability beyond latency alone. The interface also presented environmental impact using multiple equivalence formats (e.g., PET bottles, gasoline car travel, and light bulb usage), which may have introduced variation in participants’ interpretation of the magnitude of savings. In addition, tolerance for latency varied by interaction context, suggesting that fixed delays are unlikely to suit all tasks. Future eco-feedback systems should support finer-grained user control and adapt latency based on task type to better align sustainability goals with practical use.

\section{Conclusion}
In this paper, we designed and evaluated an eco-feedback LLM interface that uses service latency to reduce digital carbon emissions. Through a study with 89 participants, we examined how latency, eco-feedback design, and personal environmental norms shape eco-mode preference, usability, and anticipated adoption. Our results show that eco-feedback can support lower-carbon LLM use only within usability-compatible bounds: clear and concrete visualizations increased engagement and initial eco-mode preference, while adoption declined rapidly as response latency increased. Personal environmental norms also played an important but limited role, supporting eco-mode preference under low-latency conditions but weakening as delays and interaction frictions became more salient. Qualitative findings further suggest that eco-feedback can support learning by surfacing gaps in users’ understanding of how everyday LLM interactions relate to environmental impact and by prompting reflection during use. Based on these findings, we propose design implications for interfaces and backend systems that balance environmental signaling with responsiveness and interpretability. This work contributes to broader efforts to design sustainable, socially responsible, and learning-oriented AI systems.

\section{Acknowledgments}
We are grateful for in-kind support from Google and Amazon. This research was partially supported by the National Science Foundation under awards numbers CCF-2324860, CNS-2326608, and CCF-2324861. We would also like to acknowledge support from the Alfred P. Sloan Foundation, VMware, Google, and Catherine M. and James E. Allchin. Any opinions, findings, conclusions, or recommendations expressed in this material are those of the authors and do not necessarily reflect the views of the National Science Foundation or other supporters.

\bibliographystyle{IEEEtran}
\bibliography{ref}

\appendices
\section{User Study Materials}
\label{appendix:user_study_materials}

This appendix summarizes the main survey materials used in the study. The survey included Likert-scale, ranking, and open-ended questions.

\subsection{Eco-Mode Preference and Trade-Off Questions}
For each eco mode (Eco Modes 1--5), participants compared the eco mode with the default mode and responded to the following questions.

\begin{itemize}
    \item \textbf{Preference:} Which experience do you prefer?
    \begin{itemize}
        \item Default mode
        \item Eco mode X
    \end{itemize}

    \item \textbf{Perceived difference:} How different did you find eco mode X compared to the default mode?

    \item \textbf{Trade-off evaluation:} Participants rated how much carbon savings, delayed response, and perceived performance differences affected their preference.

    \item \textbf{Satisfaction:} How satisfied were you with your experience of eco mode X?
\end{itemize}

\subsection{Eco-Feedback Visualization Evaluation}
Participants then evaluated the eco-feedback information shown on the right side of the interface, including descriptive metrics, \textit{You Saved}, \textit{Your Impact}, and the \textit{Eco Mode Tradeoffs} graph.

\begin{itemize}
    \item \textbf{General attention:} Did you notice and pay attention to the information displayed while using the interface?

    \item For each visualization, participants reported:
    \begin{itemize}
        \item what they understood the visualization was communicating,
        \item how satisfied they were with it, and
        \item how much it influenced their experience.
    \end{itemize}

    \item \textbf{Ranking:} Participants ranked which of the following had the biggest impact on their experience:
    \begin{itemize}
        \item Descriptive metrics for each response
        \item \textit{You Saved}
        \item \textit{Your Impact}
        \item \textit{Eco Mode Tradeoffs} graph
    \end{itemize}
\end{itemize}

\subsection{Future Use and Reflection}
The survey also included questions about future willingness to use the eco-feedback interface and open-ended reflections on usability and interpretation.

\begin{itemize}
    \item If you could use this eco-feedback LLM daily, how likely would you be to use this service?

    \item Was anything confusing or hard to use? If so, please explain.
\end{itemize}

\subsection{Personal Environmental Norms}
Participants also completed a validated 10-item Likert-scale instrument measuring personal environmental norms and everyday sustainability-related attitudes and actions.

\subsection{LLM Familiarity}
Finally, the survey included background questions about prior familiarity with LLMs.

\begin{itemize}
    \item How familiar are you with Large Language Models (LLMs) like ChatGPT, Claude, Gemini, etc.?
    \item How often do you use LLM tools?
    \item What do you mostly use LLMs for? (Select all that apply.)
\end{itemize}

\subsection{Code Availability}
Additional implementation materials, including code, can be made available upon request.



\end{document}